\documentclass{article}  
\usepackage{authblk}

\usepackage{amsmath,amsfonts,amssymb}
\usepackage{graphicx}
\usepackage{hyperref}
\usepackage{siunitx}

\usepackage[usenames,dvipsnames]{xcolor}

\begin{document}                  



\title{Texture tomography, a versatile framework to study crystalline texture in 3D}
\author[1]{M.P.K. Frewein}
\author[2]{J.K. Mason}
\author[3]{B. Maier}
\author[3]{H. Cölfen}
\author[4]{A.A. Medjahed}
\author[4]{M. Burghammer}
\author[1]{M. Allain}
\author[1]{T. A. Grünewald}
\affil[1]{Aix Marseille Univ, CNRS, Centrale Med, Institut Fresnel, Marseille, France}
\affil[2]{University of California, Davis, California}
\affil[3]{University of Konstanz, Konstanz, Germany}
\affil[4]{European Synchroton Radiation Facility, Grenoble, France}









\maketitle                        


\begin{abstract}

The crystallographic texture is a key organization feature of many technical and biological materials.  In these materials, especially hierarchically structured ones, the preferential alignment of the nano constituents is heavily influencing the macroscopic behaviour of the material.
In order to study local crystallographic texture with both high spatial and angular resolution, we developed Texture tomography (TexTOM). This approach allows to model the diffraction data of polycrystalline materials by using the full reciprocal space of the ensemble of crystals and describe the texture in each voxel via a orientation distribution function. 
This means, it provides 3D reconstructions of the local texture by measuring the probabilities of all crystal orientations.
The TexTOM approach addresses limitations associated with existing models: 
It correlates the intensities from several Bragg reflections, thus reduces ambiguities resulting from symmetry. Further, it yields quantitative probability distributions of local real space crystal orientations without further assumptions on the sample structure. Finally, its efficient mathematical formulation enables reconstructions faster than the time-scale of the experiment.
In this manuscript, we present the mathematical model, the inversion strategy and its current experimental implementation. We show characterizations of simulated data as well as experimental data obtained from a synthetic, inorganic model sample, the silica-witherite biomorph. 
In conclusion, TexTOM provides a versatile framework to reconstruct 3D quantitative texture information for polycrystalline samples. In this way, it opens the door for unprecedented insights into the nanostructural makeup of natural and technical materials.

\end{abstract}


\section{Introduction}

The properties of many materials rely on their arrangement on the 
nano- and crystal structural level. 
While this organization has 
shown to be of great importance for a wide host of both technical 
and biological materials such as concrete, steel or bone, wood, 
shells and tendons to give but a few examples, its actual 
characterization poses a problem up to the current day. 
The challenge for a successful characterization during in-situ 
and in-operando studies is to enable a high spatial and angular 
resolution whilst maintaining a large field of view and ideally 
providing a non-destructive imaging modality.
While electron microscopy based techniques can boast 
impressive spatial resolution and with focused-ion beam 
tomography supplies the possibility of 3D characterization, 
the investigations are restricted to destructive 
sampling and in most cases in-vacuum operation. 
X-rays however lend themselves to the task as they easily 
penetrate millimeters, even centimeters in the case of hard X-
rays for most technical materials. Recent advances in 
nanofocusing \cite{niese_full-field_2014} has enabled the routine operation with beam 
sizes of 50 nm or less. The advent of 4\textsuperscript{th} 
generation synchrotron sources such as MaxIV \cite{tavares2018commissioning} or ESRF-EBS 
\cite{raimondi_extremely_2023} has further unlocked the potential for in-situ studies due 
to the impressive boost in flux that these machines deliver.

X-ray-based tomography methods have a long history of successful 
materials characterization. Following the routine implementation 
of full-field X-ray tomography and phase tomography \cite{paganin_x-ray_2021}, X-ray 
holo tomography \cite{cloetens_holotomography_1999} and ptychographic tomography 
\cite{dierolf_ptychographic_2010} have enabled spatial resolution on a single digit nanometer scale.
The coupling of tomography of X-ray diffraction further opened the possibility
to obtain orientation information, which is of high interest for polycrystalline materials with a non-random orientation distribution of crystallites. This property is called the \textit{crystallographic texture} \cite{kocks_texture_2000}. 
A whole class of 3D techniques exists to obtain local orientations of crystals
from the position of Bragg-peaks such as X-ray Bragg 
ptychography \cite{godard_three-dimensional_2011}, 
Bragg CDI \cite{williams_three-dimensional_2003}, 
3D-XRD \cite{poulsen_three-dimensional_2001} or DFXM \cite{simons_dark-field_2015}. 
All these techniques boast impressive angular resolution given, however, that the 
diffraction patterns show clearly distinguishable Bragg reflections.
Hierarchically structured materials such as biominerals (e.g. bone, tendon, 
shell, cuticle) but also technical ceramics and some deformed metals
can be composed of a high number of crystallites whose crystal axes are locally
distributed around a common mean orientation. 
This leads to azimuthally overlapping
diffraction peaks and yields images resembling powder diffraction patterns with
azimuthal variations in intensity, which are no longer possible to describe by a
model that specifically addresses each individual crystallite or grain. 

Another way to approach the problem is to consider first scattering tomography without orientation information, e.g. diffraction tomography approaches \cite{stock_high_2008,bleuet_probing_2008} or SAXS
tomography \cite{schroer_mapping_2006}. All these approaches assume a random orientation of the sample crystalline or nanostructural feature. In order to reconstruct directions based on nanostructural orientation required new ways to probe the reciprocal space and a parametrization for the orientation information, first shown by Georgiadis et al. \cite{georgiadis_3d_2015} for serial 2D slices
This lead to the development of tensor tomography, for which the seminal papers 
of Liebi et al \cite{liebi_nanostructure_2015,schaff_six-dimensional_2015,liebi_small-angle_2018} have shown approaches to reconstruct 
orientation tensors from the small-angle scattering signal of 
nanostructures (SASTT), which have since found a wide host of 
scientific applications. 
The technique has also been extended into the wide-angle regime
by Grünewald et al. for investigating Bragg peaks \cite{grunewald_mapping_2020} 
and used to study material properties of cartilage \cite{murer_quantifying_2021}, 
tendon \cite{silva_barreto_micro-_2024}, nervous tissue \cite{georgiadis_nanostructure-specific_2021} and bone 
\cite{grunewald_bone_2023}, to give a non-exhaustive overview. 
Recent developments in the field have seen the introduction of more flexible 
reconstruction approaches that aim at overcoming some of the 
isotropy and sampling assumptions of the original approach as well as improving reconstruction speed \cite{gao_high-speed_2019,nielsen_small-angle_2023}. In particular Nielsen et al. have presented significant performance improvements as well as an enhanced model flexibility by introducing band-limited spherical functions for the reconstruction.

All the aforementioned methods aim to reconstruct features in the diffraction pattern
by modelling their intensity directly in reciprocal space. 
By reconstructing the position of a Bragg-peak in 3D reciprocal space, for example, they 
obtain the
preferred direction of a lattice plane or axis of a crystal. This leaves, however,
one rotational degree of freedom to the orientation of the crystal, which could be 
obtained by independent reconstructions on several peaks and subsequent relation 
to the full crystal orientation tensor, in analogy to \cite{johannes_determination_2020}. 
The approach neglects however the interconnection between the positions
of the Bragg-peaks during the reconstruction, which is given by the crystal symmetry,
which in many cases is a priori well known.
Motivated by classical texture analysis, a full implementation of the orientation distribution function (ODF)  \cite{bunge_texture_1982} is used to tackle this problem. 
This means, we directly model the real space texture by assigning a probability to 
each unambiguous orientation of the crystal. The ODF fully determines the position
of all Bragg peaks at once, allowing the reconstruction of full diffraction patterns
with a single model.

Orientations of 3D objects have three degrees
of freedom, which are traditionally described by Euler angles. A framework
using a series expansion of generalized spherical harmonics \cite{bunge_orientation_1969, roe_description_1965} is often used to build the ODF and finds it use in state-of-the art software packages \cite{hielscher_novel_2008}.
The use of spherical harmonics ensures a low
number of adjustable parameters while providing the flexibility to model probabilities
all possible orientations.
Other approaches building on alternative ODF implementations and direct inversion strategies \cite{bachmann2010texture, lutterotti1997combined, matthies_reproduction_1982} exist, but are geared more towards sharp textures usually obtained in metals and geological samples and are not reviewed here in detail.

There are however shortcomings to the Euler angle parametrization, such as the 
degeneracy of orientations \cite{bunge2013texture,robinson1958use,wigner2012group}, 
a distortion of the metric tensor \cite{morawiec1990rotation} and singularities 
in the equations of motion \cite{robinson1958use}. To overcome these problems, we use an 3D harmonic expansion, which uses an axis-angle rotation 
parametrization to describe orientations\cite{mason_hyperspherical_2008}.
This framework of hyperspherical harmonics (HSH) offers additional advantages such 
as computationally efficient rotation operations and the possibility of symmetrization according to the proper point group, which drastically reduces the number of parameters 
and opens the possibility to use only the fundamental zone of the orientation space.

Coupling this versatile model for describing crystallographic textures to 
tensor tomography, we are presenting texture tomography (TexTOM) as a 
computationally efficient framework to reconstruct full 
crystallographic texture information in 3D based on scanning X-ray 
diffraction patterns. This lays the foundation for quantifying local
texture by ODFs using the full information available in the diffraction
patterns simultaneously.
The outlined method will be described in details in 
terms of its mathematical underpinning and the actual 
implementation of the code in python. Numerical simulations to 
benchmark the performance of the method will be shown alongside 
the first experimental characterization of a helicoidal 
silica-witherite biomorph as an 
example of a hierarchically textured material. 

\section{Materials and Methods}

\subsection{Biomorph sample}

Silica-carbonate biomorphs are a hierarchically structured, polycrystalline material. They are generated by the precipitation of barium carbonate in silicate-rich media at elevated pH \cite{garcia-ruiz_formation_1985,noorduin_rationally_2013}. During the formation crystalline witherite (BaCO$_3$) nanorods are formed and embedded in a silica matrix. Together, the biomorphs can take a variety of complex, curved shapes. The exact process governing the final shape is not fully understood yet\cite{kellermeier_growth_2012}. By modifying the properties of either the crystalline fraction \cite{holtus_shape-preserving_2018} or functionalizing the silica matrix \cite{helmbrecht_directed_2020,opel_functionalisation_2016,opel_light-switchable_2020} new functionalities can be added, making biomorphs an attractive material system from a material chemistry point of view.   
The biomorph sample employed here was produced a one-pot co-precipitation method\cite{opel_probing_2015}. For this, in each field of a 6-field well plate 5 mL of a 10 mM barium chloride solution were added to 5 mL of a 16.8 mM sodium meta silicate solution. The starting pH was adjusted to 11 using 1 mM NaOH. The gradual diffusion of atmospheric CO$_2$ into the solution then yielded silica-witherite biomorphs of varying shapes at the bottom of the well-plate. After 15 h the residual solution was removed and the structures were subsequently washed with water and ethanol before they were carefully detached from the bottom of the wells using a silicone brush. After transferring the biomorphs to a centrifuge vial using a pipette they were sedimented in a centrifuge using 9000 rpm for 10 min. After decanting the supernatant the structures were dried. For the synchrotron experiments, a $\sim$ \SI{60}{\micro\meter} long piece was mounted on a thin ($\sim$ \SI{10}{\micro\meter}) glass capillary with epoxy glue.

\subsection{Synchrotron experiments}

For the experimental characterization, X-ray diffraction 
experiments were carried out at the ESRF-EBS beamline ID13 
EH3 nanobranch. A 15.2 keV X-ray beam was produced by a channel-cut Si(111) monochromator and pre-focused by a set of compound 
beryllium lenses onto the final focusing optics, a set of multi-layer 
Laue lenses (MLL), producing a beam of 300x300 nm with a 
flux of 10$^{12}$~photons/s on the sample position. The sample was 
mounted on a custom-designed goniometer \cite{grunewald_mapping_2020} based on 
Smaract actuators and scanned by a piezo stage. The diffraction 
signal was recorded with an Eiger X 4M and 157.78 mm sample-
detector distance. The primary beam was blocked by a \SI{500}{\micro\meter} lead 
beamstop. The setup gave access to a usable $q$-range of 0.5 – 32 
nm$^{-1}$ and detector edges extending up to 40 nm$^{-1}$. For each projection, the full sample was scanned with a 
step-size of 500 nm in a continuous scanning mode and an adaptive 
field of view that enabled to catch the full sample (maximum size 
90 x \SI{70}{\micro\meter}) while avoiding excessive air regions around the 
sample. Diffraction patterns were recorded with 2 ms exposure 
time. Subsequently, the sample was rotated around the $z'$ axis 
and tilted around the $y'$ axis. A total of 260 projections were 
collected for 10 tilt angles between 0 and 45°. At the 0° tilt 
angle, rotation angles were collected between 0 and 180°, for all 
other tilt angles between 0 and 360°. The number of rotation 
angles in every tilt angle was reduced with a factor $\cos{\varkappa}$. 
In total, the data acquisition took 6 hours (with motor movement 
overheads). The total time of data acquisition was 3.6 h 
(6.5 Mio. diffraction patterns $\times$ 0.002 s) hours
The dose deposited on the sample was calculated according to 
Howells et al. \cite{howells_assessment_2009} as follows:
\begin{equation}
    d = \frac{\mu N_0 \epsilon}{\rho}
\end{equation}
Where $\mu = 269.58$~cm$^{-1}$ is the linear absorption coefficient for 
BaCO$_3$ at 15.2 keV, $N_0 = 2.3040 \times 10^{-15}$ m$^2$ is the incident flux per unit area, $\epsilon = 2.4370 \times 10^{-15}$~J is the photon energy and $\rho = 4.3$~g cm$^{-3}$ the mass 
density. Thus, the dose imparted on the sample 
during the full scan is $3.4 \times 10^{10}$~J/kg. Under the assumption that 
each voxel absorbs an equal amount of radiation, this equates to 
a dose of 1.3 x $10^{6}$ J/kg per voxel. 

\subsection{Simulated sample}
To test the overall functioning of the analysis, we generated data from a sample of 20$\times$20$\times$20 voxels and placed a Gaussian ODF in each voxel. That is, we defined a mean orientation $g_\mu$ and calculated the probability for each orientation the angular distance $dg$ (see appendix Eq.~\ref{eq:dg}) from $g_\mu$:
\begin{equation}
    \rho(g) = \frac{1}{\mathcal{N}} \exp\left(-\frac{dg(g,g_\mu)^2}{2\sigma^2}\right)
\end{equation}
The normalization $\mathcal{N}$ was adjusted so that $\int \rho(g) d\Omega = 1$ (volume element $d\Omega$ defined in Eq.~\ref{eq:dOmega}). Standard deviations $\sigma$ were set to 40°. These distributions were then mimicked by HSHs of order $n=12$ to facilitate rotations in the laboratory frame. The sample was generated with stripes of equal distributions in $x$-direction and random orientations along the other axes. We generated images according to Eq.~\ref{eq:full} with a BaCO$_3$ structure factor and produced 108 projections for 4 tilt angles from 0 to 45° and rotation angles in an equidistant matter as described above. Data was renormalized to a maximum number of 200 counts per bin, then Poisson noise was added to simulate the conditions of a typical measurement.

\subsection{Texture tomography inversion}

The data was regrouped into 120 azimuthal and 50 $q$ 
bins over a range from 10 to 35 nm$^{-1}$ using the PyFAI package \cite{ashiotis_fast_2015}. 
$q$-regions between the Bragg-peaks were masked before refinement.
To correct for deviations from the true rotation center of the sample,
the individual projections  were aligned using the tomographic 
self-consistency method \cite{guizar-sicairos_quantitative_2015}. 
The regular grid of 140x140x180 voxels was constructed, of which 26425 voxels were
identified as sample based on the scattering intensity in the SAXS region ($q$ 0.5 - 1 nm$^{-1}$). 
Gaussian beam intensities were calculated as 
described in sec. \ref{sec:beam} and voxels receiving $<1 \%$ of the maximum
intensity were excluded from the simulation of the respective diffraction pattern. 
BaCO$_3$ single crystal diffraction patterns were simulated
using a published witherite crystal structure \cite{holl_compression_2000}, using the lattice parameters $a = 5.3072$ \AA, $b = 8.8928$ \AA, $c = 6.4245$ \AA~ and space group Pmcn. 
Rotation symmetry generators for fundamental zone and HSH symmetrization were $2_{001}$ and $2_{010}$ as for the 222-point group.

All reconstructions were carried out on a standard compute node, 
equipped with a double CPU setup (2x AMD Epyc 7662 64 core) and 2 TB 
of RAM. The TexTOM reconstruction code is written in python, 
using numpy and numba for just in time compilation and 
parallelization of the essential parts of the code. No further code optimization has been carried out and we expect that GPU portation of the code will enable a further massive speed-up of the computations, owing to the small memory footprint of the actual inversion problem.
Further information on the reconstruction times are given in Table \ref{tab:timing}.
A damping factor of $k=2$ was used to ensure positivity of the ODF.

\section{Texture tomography}
    \subsection{Overview}
The general idea of Texture tomography is to provide a 
reconstruction scheme to extract quantitative, local crystallographic 
texture information in 3D from a 
series of X-ray diffraction patterns of a sample containing 
polycrystalline domains in various orientations. 

\begin{figure}[h!]
\includegraphics[width=\textwidth]{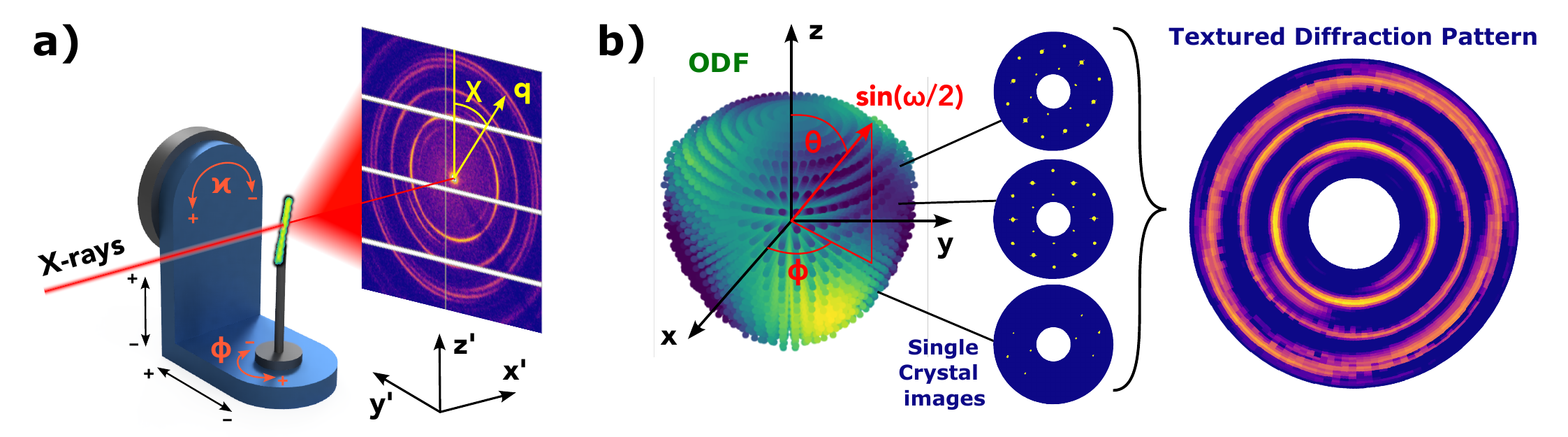}
\caption{Schematics of acquiring experimental and simulated diffraction patterns. a) The sample is raster-scanned using a focused X-ray beam in the $y'/z'$ direction for various rotation ($\phi$) and  tilt ($\varkappa$) angles. At each point, a full diffraction pattern is collected, parametrized by the momentum transfer $q$ and the azimuthal component $\chi$ of diffraction. 
b) A simulated diffraction pattern originates from an ODF and a crystal structure. The ODF is parametrized by orientations the three angles ($\omega,\theta,\phi$), which describe axis-angle rotations in the sample coordinate system ($x,y,z$). The shown ODF is color-coded so that brighter colors mean higher probability of the respective orientation. Each crystal orientation yields a different single crystal diffraction pattern and the resulting image is the sum over all of them weighted by the ODF. }
\label{fig1}
\end{figure}

A brief overview of the experimental configuration is given in 
Fig.~\ref{fig1}a. A sample is mounted on a goniometric stage that enables 
scanning ($y'$/$z'$ direction), rotation and tilting ($\phi$/$\varkappa$) of 
the sample. The sample is raster scanned with a focused X-ray 
beam and 2D diffraction patterns are collected at each scan position. This procedure is 
repeated for various tilt and rotation angles, in strict analogy 
to tensor tomography \cite{liebi_nanostructure_2015}.

The reconstructed quantity is a 3-dimensional orientation 
distribution function (ODF), representing the local arrangement of the 
crystallites via probabilities of all unambiguous orientations.
Given that the structure of a single crystal is known,
we can simulate the diffraction pattern of a polycrystalline sample
by an ODF-weighted sum over the diffraction patterns of all 
crystal orientations.
The challenge in this approach is that a faithful reconstruction
of the diffraction pattern requires a high angular resolution, to
ensure not to miss contributions from Bragg reflections from orientations
between the sampling points. Summing over the single crystal patterns
for all orientations can therefore become computationally expensive, when
it comes to parameter optimization of large samples.

We therefore choose to build up our diffraction patterns from
a basis set of elementary images, further labelled \textit{diffractlets},
which originate in a set of orthogonal functions used to model
the ODF.
The basis is given by a series expansion of 
hyperspherical harmonics (HSH)\cite{mason_hyperspherical_2008}, 
similar to a Fourier expansion in 1D and spherical harmonics in 2D.
This model therefore encodes the ODF in expansion coefficients,
which are optimized in an iterative process.
The calculation is roughly divided into the following steps:
\begin{itemize}
    \item The sample is partitioned into cubic voxels, whose 
    dimensions correspond to the stepsize of the raster scan.
    \item Each voxel contains an ODF, given by a set of coefficients of the
    HSH expansion.
    \item We calculate the projected expansion coefficients by summing
    them over all voxels, weighted by the respective beam intensity.
    \item A diffraction pattern is simulated by summing over the \textit{diffractlets} multiplied by the projected coefficients.
    \item The discrepancy between simulation and data is calculated by an error metric, which is minimized iteratively by optimizing the voxel-specific coefficients.
\end{itemize}

\subsection{Sample Translations and Rotations}\label{sec:beam}
The sample center is located at $\mathbf{x}'_0$ in the laboratory coordinate system (CS) as shown in Fig.~\ref{fig1} and we call $\mathbf{x}'_i$ the position of voxel $i$ in laboratory coordinates. We assign a central voxel around which the rotations are performed, located at $\mathbf{x}'_0$, and surround it by a cubic lattice with edge length $\Delta x$, which is the distance between two neighboring measurements. When we rotate the sample by the angles $\phi$ and $\varkappa$, we can calculate the position of voxel $i$ in the sample CS with origin at $\mathbf{x}'_0$ by:
\begin{equation}
    \mathbf{x}_i = \mathbf{x}'_0 + \hat{R}~(\mathbf{x}'_i-\mathbf{x}'_0) 
\end{equation}
using the rotation matrix for Euler angle rotations around $z'$ and $y'$
\begin{equation}
\hat{R}(\phi,\varkappa) =
        \left[ {\begin{array}{ccc}
    \cos{\phi}\cos{\varkappa} &  -\sin{\phi} & \cos{\phi}\sin{\varkappa}\\
    \sin{\phi}\cos{\varkappa} &  \cos{\phi} & \sin{\phi}\sin{\varkappa}\\
    -\sin{\varkappa}          &  0          & \cos{\varkappa}\\
  \end{array} } \right].
\end{equation}
Rotating the sample mathematically comprises the challenge of rotating all voxel positions and interpolating their values on a new grid of coordinates. This process is slow and prone to produce numerical errors. For calculating the relative positions of sample and X-ray beam, we therefore keep the sample CS static and rotate the function describing the beam by the transposed rotation matrix $\hat{R}^\intercal(\phi,\varkappa)$. In addition, we have to include translations of the sample within the $y'$-$z'$ plane (given by displacement indices $t_y$, $t_z$ and voxel size $\Delta x$). In the sample CS the translation of the beam is the negative sample displacement in the laboratory CS.

To calculate the beam path, we define 2 points traversed by the beam in sample coordinates: $\mathbf{t}$ is the point, where the beam traverses the $y'$-$z'$ plane:
\begin{equation}
    \mathbf{t}(\phi,\varkappa) = -\hat{R}^\intercal(\phi,\varkappa) \mathbf{t'} = - \Delta x~\hat{R}^\intercal(\phi,\varkappa) \left[{\begin{array}{@{\kern0pt}c@{\kern0pt}}
    0\\t_y\\t_z\end{array}}\right]
\end{equation}
$\mathbf{b}$ is another point on the beam path, resulting from adding the beam direction unit vector $\mathbf{b'} = \hat{x}'$:
\begin{equation}
    \mathbf{b}(\phi,\varkappa) = \hat{R}^\intercal(\phi,\varkappa)(\mathbf{b'} - \mathbf{t'}) = 
    \hat{R}^\intercal(\phi,\varkappa)\left(\left[ \begin{array}{@{\kern0pt}c@{\kern0pt}}1\\0\\0\end{array}
    \right] - \Delta x \left[{\begin{array}{@{\kern0pt}c@{\kern0pt}}0\\t_y\\t_z\end{array}}\right]\right)
\end{equation}

The beam intensity $B(\mathbf{x}_i)$ at each voxel position is calculated from the cumulative distribution function of the beam profile (e.g. Gaussian) in function of the normal distance $w$ between beam axis and voxel center. Note that any kind of experimentally determined beam profile can be used here. Using $\Delta x/2 - w$ as an argument gives the intensity until the voxel border, assuming that the beam width is equal or smaller than the voxel. For a Gaussian beam profile with standard deviation $\sigma$ this gives
\begin{equation}\label{eq:beam}
    B(\mathbf{x}_i) = \frac{1}{2} \biggl[ 1 - \text{erf}\left( \frac{\Delta x/2 - w(\mathbf{x}_i)}{\sqrt{2}\sigma}\right) \biggr] 
\end{equation}

The distance $w(\mathbf{x}_i)$ is calculated from the from the normal distance the line traversing points $\mathbf{t}$ and $\mathbf{b}$ and the voxel center $\mathbf{x}_i$:
\begin{equation}
    w(\mathbf{x}_i) = \frac{|(\mathbf{x}_i-\mathbf{t})\times(\mathbf{x}_i-\mathbf{b})|}{|\mathbf{b}-\mathbf{t}|}.
\end{equation}
In order to account for the orientation-dependent absorption of the diffracted beam in the sample, an implementation of the absorption correction outlined in \cite{grunewald_bone_2023} can be used. Due to the low absorption of the samples employed in this study, it was currently not implemented.

\subsection{Single Crystal Diffraction Patterns}

The diffraction pattern of a single crystal $I_{sc}(\mathbf{q})$ is calculated from all the atom positions $\mathbf{R_j}$ from the structure factor $S(\mathbf{q})$.
\begin{equation}\label{eq:sf}
    I_{sc}(\mathbf{q}) \propto S(\mathbf{q}) = \frac{1}{\sum_i f_i} \sum_j \sum_k f_j(\mathbf{q}) f_k(\mathbf{q}) \text{e}^{i\mathbf{q}(\mathbf{R_j}-\mathbf{R_k})},
\end{equation}
We use atomic form factors $f(q)$ as tabulated in \cite{henke_x-ray_1993}.

The crystal is created from a known crystal structure and the unit cell is repeated in 3 dimensions to resemble the expected crystal size. We would like to note that the interface is open to accept other input sources for $S(\mathbf{q})$ such as Discus  \cite{proffen_discus_1997}  in order to provide a more detailed modelling of the crystalline parameters in the future. 
The $\mathbf{q}$-vectors corresponding to X-ray wavelength $\lambda$ and experimental setup are calculated using the surface of the Ewald sphere \cite{grunewald_photon_2016}.
With the beam oriented along $\hat{x}'$ (see Fig.~\ref{fig1}), the origin of the sphere in reciprocal space is at $(-2\pi/\lambda,0,0)$ and its radius is $2\pi/\lambda$. We express these points in terms of the momentum exchange $q = \frac{4\pi}{\lambda}\sin{\theta}$ and the angle in the detector plane $\chi$
\begin{align}\label{eq:ewald}
    \mathbf{q} = \left( {\begin{array}{c}
        q \cos(\theta')\\
        - q \sin(\theta') \sin(\chi)\\
        q \sin(\theta') \cos(\chi)\\
    \end{array} } \right), \qquad \theta' = \pi/2 + \theta,
\end{align}
with $q$ and $\chi$ corresponding to the polar coordinates on the detector, defined in Fig.~\ref{fig1}.

In the construction of textured diffraction patterns, it is necessary to compute the contribution from each unambiguous orientation of the crystal. 
 The diffraction pattern $I^g_{sc}(q,\chi)$
of a crystal in orientation $g$ is calculated by rotating all atom positions and
using the structure factor (Eq.~\ref{eq:sf}).

\subsection{Orientation Distribution Functions}\label{ref:odf}
An orientation distribution function (ODF) $\rho(g)$ assigns a probability to a rotation $g$ from a reference orientation and
is expedient to describe properties of an ensemble of objects with different orientations. 
We use it to model the scattering intensity $I(q,\chi)$ produced 
by an ensemble of crystallites, assuming all of them have the same crystal structure in different orientations. 
The ODF connects single crystal diffraction patterns $I^g_{sc}$ to an image resulting
from an ensemble by an integration over all crystallite orientations $g$:
\begin{equation}\label{eq:int_odf}
    I(q,\chi) = \int \rho(g)~I^g_{sc}(q,\chi)~d\Omega.
\end{equation}

\subsubsection{Hyperspherical Harmonic Expansion}

We model the ODF as a series expansion of hyperspherical harmonics (HSH) $Z^n_{lm}(g)$ (see appendix and \cite{mason_thesis}).
These are complex functions that are naturally written using the axis-angle rotation parametrization.
We use these rotations to describe crystal orientations with respect to
the fixed sample CS. The axis-angle parameterization expresses a 3D rotation
using a single rotation
by an angle $\omega\in[0,\pi]$ around a unit vector axis $\mathbf{\hat{a}}$, which we
define by polar angle $\vartheta\in[0,\pi]$ and azimuthal angle $\varphi\in[0,2\pi)$:
\begin{align}
    \mathbf{\hat{a}} = \left( {\begin{array}{c}
        \sin(\vartheta) \cos(\varphi)\\
        \sin(\vartheta) \sin(\varphi)\\
        \cos(\vartheta)\\
    \end{array} } \right).
\end{align}
The volume element for this parametrization is given by:
\begin{equation}\label{eq:dOmega}
    d\Omega = \sin(\omega/2) \sin\vartheta d\omega d\vartheta d\varphi.
\end{equation}
Since the HSHs form an orthonormal basis for functions of rotations, it is possible to construct an ODF as a linear combination of HSHs, where $Z^{0}_{0,0}$ provides the probability mass and higher order HSHs redistribute this probability mass as a function of $g$. The total ODF depends on a set of complex complex coefficients $c^n_{l,m}$ as:
\begin{equation}\label{eq:odf}
    \rho(g) = \sum_{n,l,m} c^n_{lm} Z^n_{lm}(g)
\end{equation}
Furthermore, the ODF of a rotated sample can also be written in the form of Eq.~\ref{eq:odf} with a different set of coefficients. The transformation of the coefficients depends on the rotation to which the sample is subject.
This is described by the matrix $R^{n/2}_{l' m' l m}(g_l, g_r)$ \cite{mason_expressing_2009}, which allows one to write the effect of a rotation of a HSH $Z^n_{l,m}$ as a linear combination of other HSHs $Z^n_{l',m'}$ of the same order $n$ (see appendix~\ref{sec:app_hsh})
\begin{equation}\label{eq:HSHrot}
    \rho'(g_lgg_r) = \sum_{n,l,m} \sum_{l'm'} c^n_{lm} Z^n_{l'm'}(g) \,R^{n/2}_{l'm'lm}(g_l, g_r).
\end{equation}
The variable $g$ of the initial ODF defined in Eq.~\ref{eq:odf} is related to the variable $g' = g_l g g_r$ of the rotated ODF, where $g_r$ is one of the point symmetry operations of the crystal and $g_l$ is a rotation of the sample in the laboratory frame.
This rotation $g_l$ will be used to simulate sample rotations in the course of the tomography experiment. 
The other rotation $g_r$ will always be the null rotation $g_0$.

\subsubsection{Symmetrized Hyperspherical Harmonics}
HSHs, as defined in Eq.~\ref{eq:HSH}, are complex valued functions and require certain combinations of complex coefficients to produce a real-valued ODF. 
We therefore use symmetrized hyperspherical harmonics (sHSHs) $\mathring{Z}^n_\lambda(g)$, which obey this constraint by definition. The sHSHs, as functions of rotations, can also be written as linear combinations of the HSHs of the same order $n$:
\begin{equation}
    \mathring{Z}^n_\lambda(g) = \sum_{l,m} Z^n_{lm}(g) X^n_{lm;\lambda}.
\end{equation}
The symmetrization procedure to obtain $X^n_{lm;\lambda}$ is outlined in Mason \& Schuh \cite{mason_hyperspherical_2008}. 
This strategy additionally allows selecting sHSHs with the same proper point group as the crystal structure. 
This reduces the number of adjustable parameters, and the crystal orientations that need
to be evaluated are limited to the fundamental zone of the point group \cite{heinz_representation_1991}.
The matrix for the effect of a single rotation on the coefficients $\mathring{c}^n_{\lambda,k}$ of the expansion over the sHSHs is then given by:
\begin{equation}\label{eq:sHSHrot}
    \mathring{R}^n_{\lambda',\lambda}(g) = \sum_{l'm'}\sum_{l,m} X^{n\dagger}_{l'm';\lambda'} R^{n/2}_{l'm'lm}(g,g_0) X^n_{lm;\lambda}.
\end{equation}

\subsection{Full forward model}

\begin{figure}[h!]
\includegraphics[width=\textwidth]{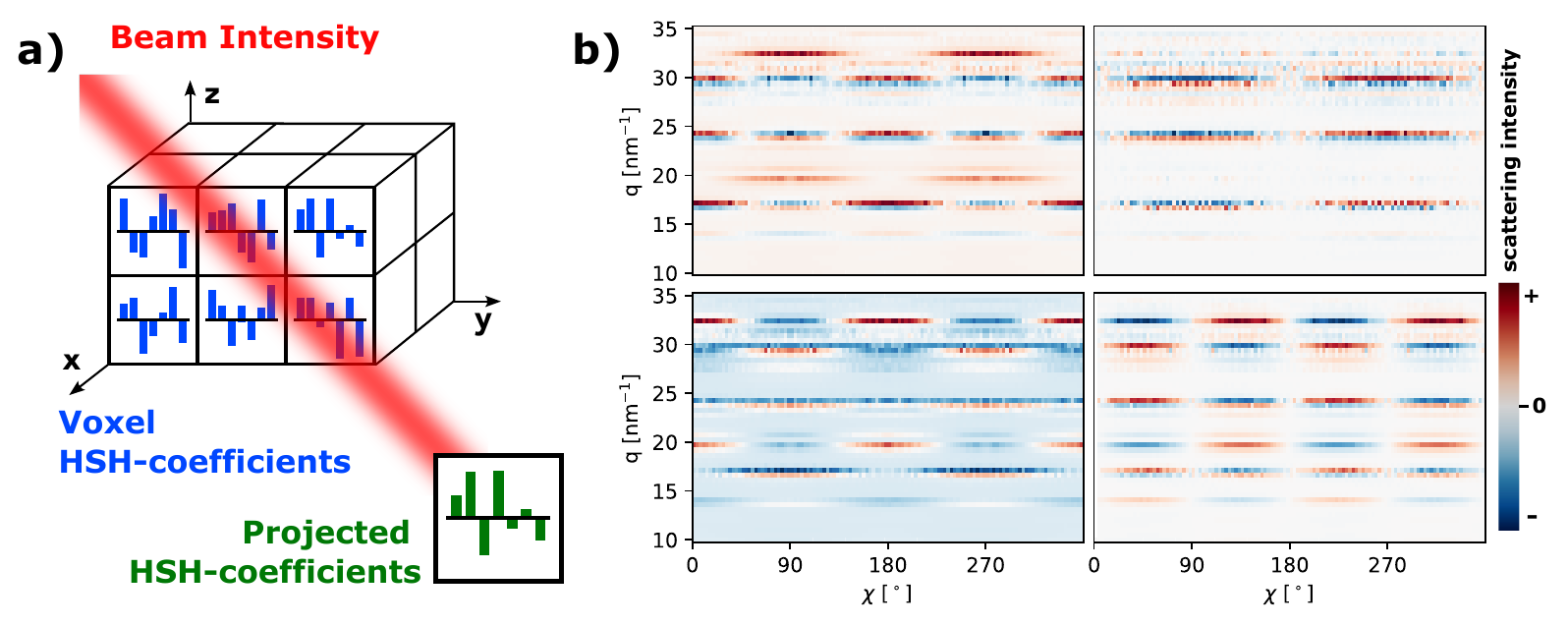}
\caption{
a) A sample with HSH expansion coefficients in each voxel, represented by bar diagrams. These are weighted by the beam intensity for a given configuration and result in a projection of these coefficients. A summation of \textit{diffractlets} weighted by these coefficients results in a diffraction pattern.
b) is a selection of \textit{diffractlets} of order 4 sHSHs with a BaCO$_3$ structure factor. 
}
\label{fig2}
\end{figure}

Let us first derive the forward model for a sample made of a single voxel, located in $\mathbf{x}_1$. We can directly use the sHSH model above to calculate the diffraction patterns expected in that case: let us expand the ODF in Eq.~\ref{eq:int_odf} using sHSHs, the expected scattering intensity in any reciprocal space coordinates $q$ and $\chi$ reads then
\begin{equation}\label{eq:I_k}
    \bar{I}(q,\chi; \mathbf{c}(\mathbf{x}_1)) =  \sum_{n,\lambda} c^n_{\lambda}(\mathbf{x}_1) \int \mathring{Z}^n_\lambda(g)~I^g_{sc}(q,\chi)~\text{d}\Omega  
\end{equation}
where $\mathbf{c}(\mathbf{x}_1) := \{ c^n_{\lambda} \,|\, (n,\lambda) \in \mathbb{N}\times \mathbb{N}\}$
is the set of sHSH coefficients associated with the single voxel. 
Here, we isolate the contribution to the diffraction pattern by a single sHSH, further labeled \textit{diffractlet} (see Fig.~\ref{fig2} b), which reads
\begin{equation}
    d^n_\lambda(q,\chi) :=  \int \mathring{Z}^n_\lambda(g)~I^g_{sc}(q,\chi)~\text{d}\Omega.
\end{equation}
Obviously, any realistic sample should be defined over a mesh, with a series of $P$ voxels located in $\{\mathbf{x}_p\}_{p=1}^P$.  
It is then natural to associate any voxel $p$ with its specific series of sHSH coefficients, hence 
resulting in a set of parameters in the model 
that reads 
\[
\mathbf{c} := \{c^n_{\lambda}(\mathbf{x}_p)\,|\, p=1,\cdots,P\} \qquad \text{with} \qquad 
(n,\lambda) \in \mathbb{N}\times \mathbb{N}.
\]
In order to account for the effect of the tomographic measurement, we introduce another index $k = 1 \cdots N_\phi\times N_\varkappa\times N_t$ that refers to a given sample rotation $\phi$, tilte
$\varkappa$ and translation $\mathbf{t}$.
The total diffraction pattern (\textit{i.e.,} the expected measurement) is a weighted sum of the contribution of each voxel in the sample, with the weights computed from the local beam intensity $B_k(\mathbf{x}_p)$, as given by Eq.~\ref{eq:beam} (illustrated in Fig.~\ref{fig2} a). For a single measurement $k$, we note that most of the weights are zero as only a small fraction of the voxels in the sample are actually illuminated by the beam. 
The sHSH coefficients $c^n_{\lambda}(\mathbf{x}_p)$ are agnostic to any rotation and tilt in the sample and we need to explicitly account for it in the forward model. To that end, we introduce first the sample rotation $g_k$ that is associated with the $k$-th tomographic measurement.  Then, we use the rotation matrix given by (Eq.~\ref{eq:sHSHrot}) to
define the corresponding \textit{rotated diffractlets} 
\[
\delta^n_{\lambda,k}(q,\chi) := \sum_{\lambda'} d^n_{\lambda'}(q,\chi)\, \mathring{R}^n_{\lambda',\lambda}(g_k).
\]
It is then straightforward to define the expected intensity for the $k$-th tomographic measurement    

\begin{align}\label{eq:full}
    \bar{I}_k(q_\ell,\chi_m; \mathbf{c}) 
    &= \sum_p \sum_{n} \sum_{\lambda}   B_k(\mathbf{x}_p) \delta^n_{\lambda,k}(q,\chi) c^n_{\lambda}(\mathbf{x}_p)
\end{align}
where $(q_\ell, \chi_m)$
define the measurement mesh in the reciprocal space coordinates. Clearly, this  relation is a linear model that we write now in a convenient matrix-vector form: 
$i$) first, we wrap the measurement-related indices $k,\ell,m$ into a single index $j = 1\cdots N_k N_q N_\chi$,  then $ii)$ we use a single index $\nu$ in place of the 3 model-related parameters $p,n,\lambda$. The relation Eq.~\ref{eq:full} now reads
\begin{equation}\label{eq:discret}
    \bar{I}_j(\mathbf{c}) = \sum_\nu B_{j,\nu} \delta_{j,\nu} c_\nu = \mathbf{a}_{j}^\dagger
    \mathbf{c}
\end{equation}
and if we stack the expected measurements 
$\bar{I}_j$ into a single vector, we obtain
\begin{equation}
\label{eq:vect}
    \bar{\mathbf{I}} =  \mathbf{A}
    \mathbf{c}
\end{equation}
where $\mathbf{A}$ is the texture-tomography matrix that can be pre-computed before we perform the iterative inversion.

\subsection{TexTOM reconstruction strategy}

We use a standard quadratic metric 
\begin{equation}
    \label{eq:LSCriterion}
    \mathcal{L}(\mathbf{c}) = ||\,\mathbf{A}
    \mathbf{c} - {\mathbf{I}} \,||^2
\end{equation}
to compute  the discrepancy between the output of the forward model (Eq.~\ref{eq:vect}) and the texture-tomographic experimental measurements $\mathbf{I} := \textbf{vect}(I_j)$. 
The  reconstruction is defined as the minimization of the following constrained least-square criterion   
\begin{equation}
    \label{eq:least-square}
    \widehat{\mathbf{c}} = \arg \min_{\mathbf{c}\in\mathbb{R}^N} 
    \mathcal{L}(\mathbf{c}) 
    \qquad \text{subject to}\qquad \{c^0_{0}(\mathbf{x}_p) \geq 0\}_{p=1}^P.
\end{equation}
We note that the positivity constraint over the zero-order sHSH coefficients is required to produce a physically meaningful results. 
This constraint can be fulfill by projecting any negative $c^0_{0}(\mathbf{x}_p)$ to zero after each update of the following gradient-base iteration 
\begin{equation}
    \label{eq:iteration}
    \qquad 
    \mathbf{c}^{(n+1)} \leftarrow \mathbf{c}^{(n)} - \gamma^{(n)} \nabla \mathcal{L}(\mathbf{c}^{(n)}), 
    \qquad n = 1 \cdots \infty.
\end{equation}
The gradient of the least-square function $\nabla \mathcal{L}$ is easily derived from 
Eq.~\ref{eq:LSCriterion}
\begin{equation}
    \nabla \mathcal{L}(\mathbf{c}) = 2 \mathbf{A}^\dagger (\mathbf{A}
    \mathbf{c} - {\mathbf{I}})
\end{equation}
The stepsize $\gamma^{(n)}$ is adjusted in each iteration with a \textit{backtracking technique} to ensure a strict decrease in the criterion \cite[p.~29]{Bertsekas99}. 
We note that the fitting-function Eq.~\ref{eq:LSCriterion} is \textit{strictly convex} whenever  $\mathbf{A}$ is a full-column rank matrix (which is expected to be the case here). 
In this context, the solution of the constrained optimization problem Eq.~\ref{eq:least-square} exist and is unique \cite[Prop.~1.1.2]{Bertsekas99}, and  the convergence of the iteration Eq.~\ref{eq:iteration} toward $\widehat{\mathbf{c}}$ is granted, whatever the initial-guess $\mathbf{c}^{(0)}$. 
The convergence of the iteration is monitored \textit{via} the norm of the gradient, and we stop the iterative reconstruction when 
\begin{equation}
\biggl|\frac{||\nabla \mathcal{L}(\mathbf{c}^{(n+1)})||-||\nabla \mathcal{L}(\mathbf{c}^{(n)})||}{||\nabla \mathcal{L}(\mathbf{c}^{(0)})||}\biggl| < \epsilon
\end{equation}
where $\epsilon$ is a predefined parameter (typically, it is set to $10^{-3}$). 

\subsection{Data post-processing}\label{sec:kernel}
A known limitation of the harmonic expansion approach is that it hard to ensure positivity of the ODF\cite{hielscher_novel_2008}, in particular without explicitly calculating the ODF in every iteration, which can be computationally extremely costly for large samples. We therefore post-process our data using a kernel that ensures non-negativity \cite{mason_convergence_2013}. This means we modify the obtained coefficients by a prefactor $K$ that depends on the order $n$, the highest used order $N$ and an exponent $k$, whose value is empirically chosen to be between 1 and 2 according to the situation.
\begin{equation}
    K = \left( 1-\frac{n}{N+1} \right)^k.
\end{equation}
It was observed however that this kernel slightly spreads out the ODF, therefore the presented standard deviations are possibly over-estimated.

\section{Results and Discussion}
\subsection{Reconstruction of simulated data}

\begin{figure}[h!]
\includegraphics[width=\textwidth]{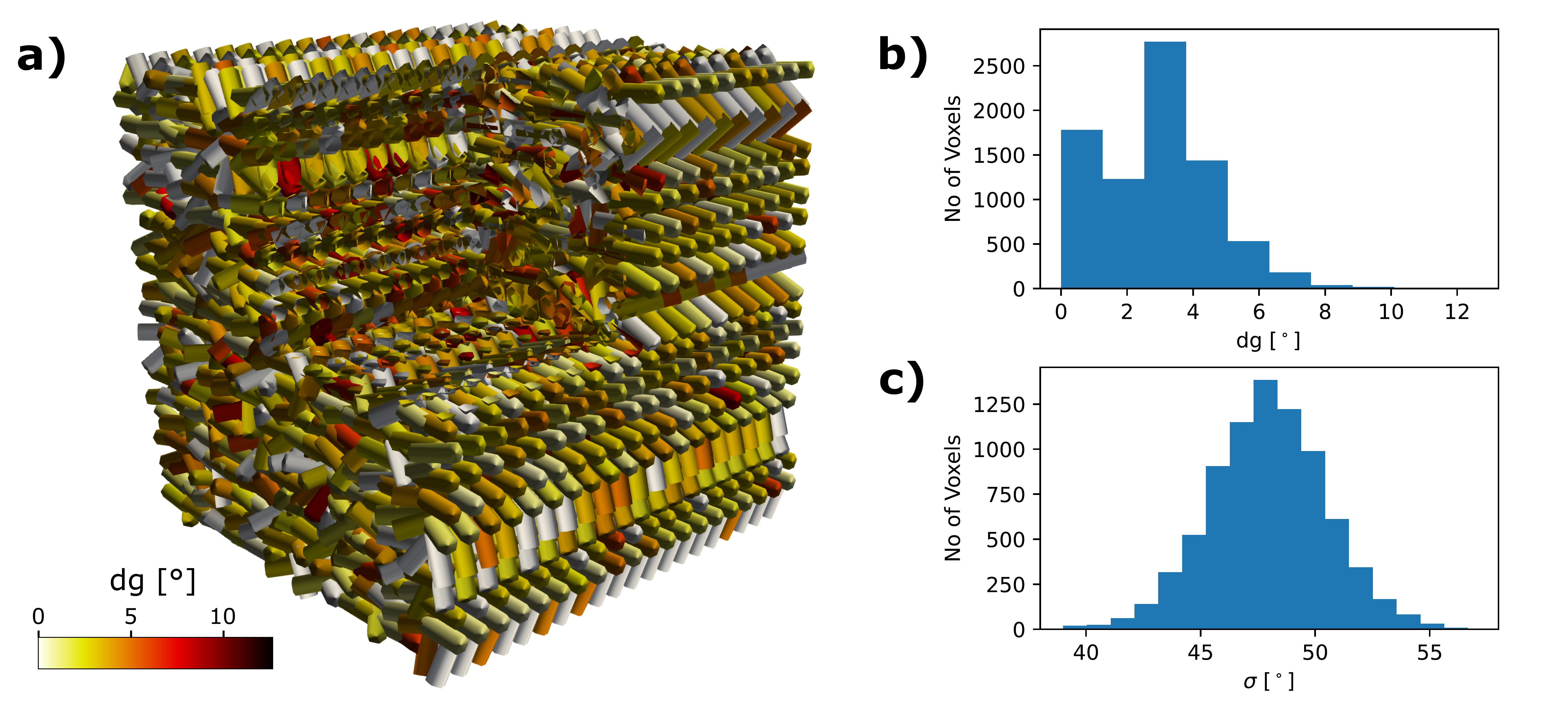}
\caption{a) TexTOM reconstruction of the simulated sample for testing the reconstruction algorithm. The sticks represent the reconstructed preferred orientation of the crystal c-axis in each voxel, color coded by the angular deviation $dg$ from the simulation. A corner was cut out for showing the interior of the sample. b) histogram of $dg$ for the same sample. The distribution of the standard deviations is shown in c).}
\label{fig_sim}
\end{figure}

The reconstruction of ODFs of the simulated sample was performed in two steps: First is the retrieval of the mean orientation using a HSH-expansion cut at the lowest order, second the estimation of the variance by including higher orders as necessary. 
Figure~\ref{fig_sim} presents a summary of the inversion results: Panel a) shows a cut-off view of the reconstruction. Note that this representation shows only the orientation of one crystal axis, thus does not represent the full texture information. 
The colour scale is the angular distance metric (see appendix \ref{sec:app_dg}) between the simulation and the reconstruction, which is shown as a histogram in panel b). The deviations are distributed around an average of 2.8° and show no clear spatial distribution at the interface between differently oriented layers of the sample. The irregular shape in this histogram is connected to the distribution of sampling points in orientation space, which was constructed with an angular resolution of 3°.
This part of the reconstruction was done with the expansion truncated at order 4, where the point group 222 possesses 10 sHSHs, and demonstrates that the lowest order already suffices for the estimating the most likely orientations. 

An expansion up to order 8 was used to estimate standard deviations $sigma$ of the ODFs. As discussed in section \ref{sec:kernel}, our current model over-estimates the spread of the distribution as a consequence of ensuring non-negativity of the ODF. We find $\sigma$ distributed around 47.9° with a standard deviation of 2.6°, as shown in panel c) of Fig \ref{fig_sim}.
\subsection{Biomorphs}
\label{Sec:biomorph}
\begin{figure}[h!]
\includegraphics[width=\textwidth]{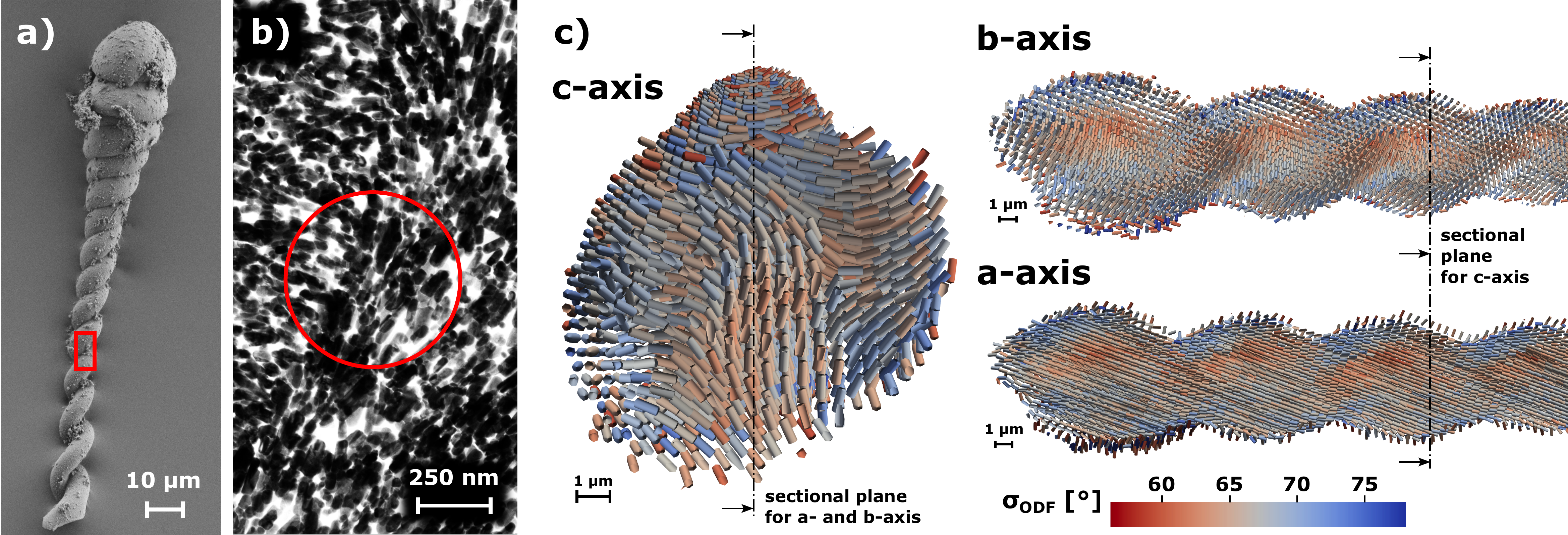}
\caption{ a) SEM image of a helicoidal silica biomorph. b) Arrangement of BaCO$_3$ nanorods (black) in amorphous silica (white) as seen by TEM. For reference, size of the shown region is given by the red rectangle in a). The red circle corresponds to the dimension of the X-ray beam. c) TexTOM reconstructions of a \SI{60}{\micro\meter} long piece of a helix. Sticks represent the preferred orientation of the indicated crystal axes. The volume is cut in transverse (c-axis) and longitudinal (a- and b-axes) directions, thus showing the interior of the sample. Images were produced by Paraview \cite{ParaView}}
\label{fig3}
\end{figure}

In order to test the performance of TexTOM on experimental data, a dataset of silica biomorph was collected. The sample consisted of a helicoidal silica-witherite biomorph of \SI{60}{\micro\meter} length and \SI{15}{\micro\meter} diameter. An exemplary SEM image can be seen in Fig.~\ref{fig3}a.  The arrangement of  BaCO$_3$ nanorods of about 20$\times$100~nm embedded in an amorphous silica matrix is shown in the TEM-image, Fig.~\ref{fig3}b. Different morphologies can be obtained by varying the local synthesis conditions \cite{noorduin_rationally_2013,kellermeier_growth_2012}. In the context of this manuscript, no in-depth analysis of different growth conditions or their texture has been carried out. Shown data server as a sample to demonstrate the performance of TexTOM.

\begin{figure}[h!]
\includegraphics[width=0.5\textwidth]{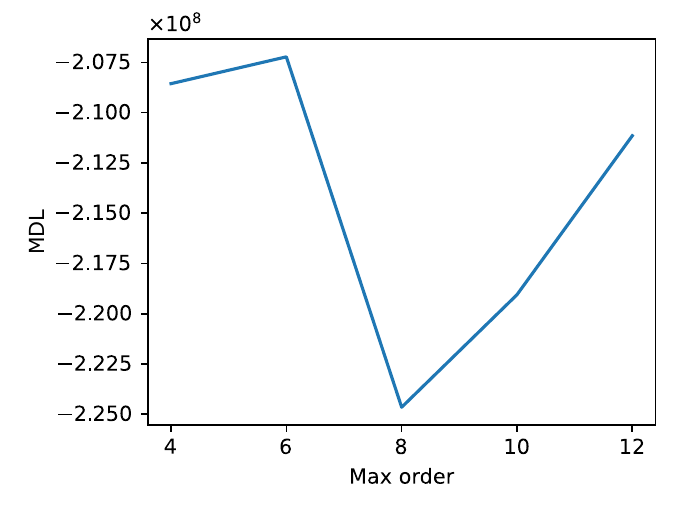}
\caption{ Plot of the minimum description length (MDL) for the reconstruction of the helicoidal biomorph in function of the HSH order $n$.}
\label{fig_order}
\end{figure}

A dataset containing 260 projections was collected, with an equal angular sampling in $\phi$ and $\varkappa$ based on the reasoning of Liebi et al. \cite{liebi_small-angle_2018} to create a 'gold standard' measurement. From the reconstructions, the most likely orientation of the a, b and c-axis was determined within in the fundamental zone and the variance was extracted as a metric for the degree of orientation/angular dispersion within each voxel and plotted as the colour scale of Fig.~\ref{fig3}c. In order to facilitate the interpretation of the data, cross sections of the volume are presented in Fig.~\ref{fig3}c. The reconstructions show an angular dispersion consistent with TEM observations. The red circle indicated the beam size in comparison to the TEM image. In the slice of the c-axis the two distinct strands of the helix can be identified. An interesting observation is the presence of a gradient of the variance along the two helix strands, visible in the cuts of the sample showing a and b-axes of Fig.~\ref{fig3}c. These possibly correspond to a core-skin architecture in the arrangement of the nanorods. This could be caused by local concentration and pH gradients during the synthesis. A more detailed study, encompassing an in-depth, comparative analysis of different morphologies is envisaged for the future. 
The reconstruction strategy (described in more detail in the Materials and Methods section) is split into two major parts, one is the precalculation of single crystal diffraction patterns and the beam trajectories corresponding to the scan. The rational here is that these time consuming steps can be re-used for different reconstructions. 
The actual reconstructions are carried out with sequentially increasing order, which 
allows the selection of an optimal order via a \textit{Minimum Description Length} 
(MDL) criterion \cite{Hansen01}. In our case, the MDL criterion is clearly selecting  
the order 8 (see Fig.~\ref{fig_order}). A more detailed account of the timing is given in Tab.~\ref{tab:timing}. 

\begin{table}[]
    \caption{Duration $t$ of different steps of TexTOM reconstructions and total number of adjustable parameters $N_{par}$.}
    \vspace{10pt}
    \centering
    \begin{tabular}{l|rr}
        \textbf{Pre-calculations}& t [min]\\
         Single crystal diffraction patterns & 770 \\
         Beam trajectories & 5\\
         \textbf{260 projections}& t [min] & $N_{par}$\\
         \hline
         Reconstruction $n=4$ & 35 & 290675 \\
         Reconstruction up to $n=8$ & 100 & 1189125\\
         Reconstruction up to $n=12$ & 305 & 3144575\\

         \textbf{41 projections}&t [min] & $N_{par}$\\
         \hline
         Reconstruction $n=4$ & 5& 290675 \\
         Reconstruction up to $n=8$ & 24 & 1189125\\
         Reconstruction up to $n=12$ & 60 & 3144575\\
    \end{tabular}
    \label{tab:timing}
\end{table}

\subsection{Benchmarking}
\begin{figure}[h!]
\includegraphics[width=\textwidth]{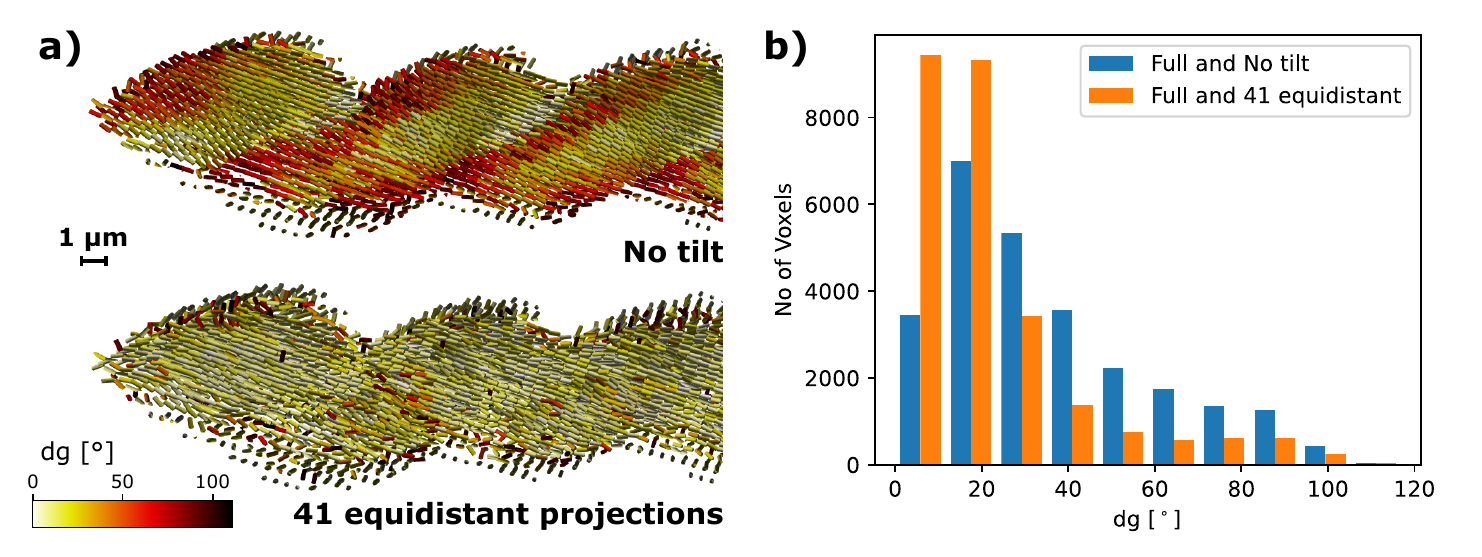}
\caption{Comparison of reconstructions using all 260 projections and either 41 projections with $\kappa=0$ or 41 projections with an equidistant distribution over orientation space. a) shows how the angular deviations $dg$ from the full reconstruction are distributed over the interior of the sample, using the same sectional plane as in Fig.~\ref{fig3}. b) shows the distribution of $dg$ for both cases.}
\label{fig_bench}
\end{figure}

In order to test the performance of our code and see how under-sampling affected our results, two further datasets were selected from the existing data. One being all 41 projections at $\varkappa$=0 and another being 41 projections, equally sampled in $\phi$  and $\varkappa$ orientation space. The sampling with 41 projections corresponds roughly to a Nyquist-Shannon sampling \cite{shannon1949}. Reconstructions were carried out and compared by calculating the angular distance between the most likely orientations for each voxel, shown in Fig.~\ref{fig_sim}a.
The histogram in Fig.~\ref{fig_bench}b shows amount of voxels for each distance. It is visible that the 41 equidistant projections reproduce the orientation very well with a mean angular distance of 22° and a decrease in quality that can be expected from the sampling reduction. Here, there are no visible regions with a worse reconstruction quality, it rather adds a general noise on the results.
A similar, but slightly worse trend is visible for the $\varkappa$=0 dataset. Here, the mean angular distance is 35° and the histogram (blue bars, Fig~\ref{fig_bench}b) shows a higher tail at larger angles. Furthermore, the large orientation differences seem to be linked to regions where the crystallites have a wide ODF (compare red zones in top panel of Fig.~\ref{fig_bench} a and blue zones in Fig.~\ref{fig3}) hence, a weak texture, whereas voxels with a strong texture are in good agreement. 
Together with the very fast reconstruction time (5 min) for n=4 for the reduced datasets, the strategy of online reconstruction and information-driven sampling comes into reach. This strategy entails that continuous online reconstructions are carried out and that projections are added in zones of the reciprocal space where the fit shows larger deviations between reconstructed and measured data. 

\section{Conclusions and Outlook}
In summary, this manuscript presents TexTOM as a new inversion framework to recover quantitative ODF information in a tomographic fashion from X-ray diffraction data using hyperspherical harmonics and derived \textit{diffractlets}. We present a detailed description of the experiment, the forward model including the parametrization of the HSH expansion as well as our inversion strategy.
We show the results of an inversion on both simulated and experimental data with sub $\mu$m resolution and benchmark the reconstruction strategy with reduced angular sampling. 

The presented method presents a large step forward from state-of-the-art tensor tomography methods \cite{liebi_nanostructure_2015, gao_high-speed_2019, nielsen_small-angle_2023} by enabling a fully quantitative reconstruction of the real-space ODF compared to the reconstruction of the position of a single SAXS or WAXS reflection.
Furthermore TexTOM does not rely on regularization parameters and associated assumptions on local smoothness, but uses the inherent constraints given by the crystal symmetry. 
Due to the nature of the harmonic decomposition, the recovery of mean orientations is given by the lowest order and therefore allows very fast reconstructions, even when compared to the actual data acquisition. 
The estimation of the spread of the ODF requires the inclusion of higher orders and  the use of a damping kernel also enables to enforce strict positivity of each component of the ODF. This method currently overestimates the true variance and and we envisage further tuning if the model to retrieve better estimates.

The joint-optimization of several Bragg-peaks also enables the concurrent refinement of multiple crystalline phases within one inversion as well as adding a strain component. For the future, the direct extraction of crystalline phase information prior to a TexTOM inversion via e.g. a Pawley extraction can also be envisaged, requiring less prior information on the crystalline phases present. 
In this way, TexTOM aims at reconstructing the full crystalline state tensors for each voxel with high, essentially beam-size limited resolution. One challenge arising from the wealth of information contained in the ODF is to find ways to visualize the retrieved data. Further challenges are arising when trying to push the experimental resolution down into the range of 100~nm or less. Firstly, the data acquisition becomes more challenging as both the scanning and rotation need to provide a positioning accuracy better than the target resolution. Secondly, the alignment of the data during the pre-processing is subject to equal constraints and factors such as imperfections in the tilt and rotation axis as well as the coaxiality of the scanning stages becomes more and more important and might approach the limit of current mechanical solutions. One way to overcome this challenge could be the use of nonrigid tomography approaches \cite{odstrcil_ab_2019}

Through the efficient use of the collected diffraction information, a significant speed-up of the experiment can be expected, bringing in-situ experiments on dynamically changing samples into reach as well as providing an avenue for measuring radiation sensitive samples with a largely reduced deposited X-ray dose. 
Future work will be directed towards accurately benchmarking the performance in terms of spatial, angular and multi-phase resolution with specifically designed benchmark samples. A further of development is implementation of live reconstructions and information-driven sampling to realize the quickest-possible experiments whilst achieving the desired angular and spatial resolution.


\appendix
\section{Equations related to hyperspherical harmonics}\label{sec:app_hsh}
Definition of hyperspherical harmonics \cite{mason_thesis}:
\begin{align}\label{eq:HSH}
    \begin{split}
        Z^n_{lm}\Big(g(\omega,\vartheta,\varphi)\Big) = (-i)^l \frac{2^{l+1/2}l!}{2\pi} \sqrt{(2l+1) \frac{(l-m)!(n+1)(n-l)!}{(l+m)!(n+l+1)!}}\\
        \Big[\sin(\omega/2)\Big]^l C^{l+1}_{n-l}\Big[\cos(\omega/2)\Big]P^m_l\Big(\cos\vartheta\Big)\text{e}^{im\varphi}
    \end{split}
\end{align}
The $C^{l+1}_{n-l}$ are the Gegenbauer polynomials and the $P^m_l$ are the associated Legendre polynomials with the Condon–Shortley phase.
Their integer indices are restricted to $n\geq0$, $0\geq l \geq n$, $-l\geq m\geq l$. 

The hyperspherical harmonic rotation matrix is given by:
\begin{equation}
    R^{2l}_{\lambda'\mu'\lambda\mu}(g_2, g_1) = \sum_{m_2'}\sum_{m_2}\sum_{m_1'}\sum_{m_1} C^{\lambda'\mu'}_{lm_2lm_1'} U^{l*}_{m_2'm_2}(g_2) U^{l}_{m_1'm_1}(g_1) C^{\lambda\mu}_{lm_2'lm_1}
\end{equation}
where $U^l_{m,m'}$ is the irreproducible representative of $SO(3)$
\begin{equation}
    U^l_{m,m'}(g) = \sum_\lambda \sum_\mu \frac{\sqrt{2(2\lambda+1)}\pi}{2l+1} C^{lm}_{lm'\lambda\mu} Z^{2l}_{\lambda\mu}(g)
\end{equation}
and $C^{lm}_{lm'\lambda\mu}$ being the Clebsch-Gordon coefficient for $SO(3)$.

\section{Angular distance between orientations}\label{sec:app_dg}
We define the distance between two orientations $g, g'$ via the quaternion formalism. Unit quaternions are related to rotation angle $\omega$ and axis $\mathbf{\hat{a}}$ by: 
\begin{equation}
    \mathbf{q} = q_0 +q_1\mathbf{i} +q_2\mathbf{j} +q_3\mathbf{k} = \cos\frac{\omega}{2}+\sin\frac{\omega}{2} (a_x\mathbf{i} +a_y\mathbf{j} +a_z\mathbf{k})
\end{equation}
The angular distance between two unit quaternions is then given by \cite{huynh2009metrics}
\begin{equation}\label{eq:dg}
    dg(g,g') = \arccos ( 2 \langle \mathbf{q}, \mathbf{q}' \rangle -1  )
\end{equation}
Here, $\mathbf{q}, \mathbf{q}'$ are the quaternions corresponding to rotations $g,g'$, respectively. $\langle \mathbf{q}, \mathbf{q}' \rangle$ denotes the inner product of the quaternions:
\begin{equation}
   \langle \mathbf{q}, \mathbf{q}' \rangle = q_0q'_0 +q_1q'_1 +q_2q'_2 +q_3q'_3
\end{equation}
This however, does not yet take into account the crystal symmetry. To find the effective minimal distance between two orientations, the therefore rotate $\mathbf{q}$ by the symmetry generators of the point group and choose the smallest distance of all possible combinations.

\section{Preferred orientation and standard deviation of ODFs}

We examined the resulting ODFs resulting from the sHSH-expansion and found that the ODF truncated at the lowest order ($n=4$ for point group '222') generally shows a single maximum within the fundamental zone. We define this orientation as the mean orientation $g_\mu$ of the distribution.
To calculate a standard deviation of an ODF given by a set of coefficients $\mathbf{c}$, we rotate the distribution by $-g_\mu$ using relation \ref{eq:HSHrot}, thus obtaining a distribution centered at the center of the fundamental zone (where $\omega=0$). Then we calculate the standard deviation $\sigma$ in the conventional way from the rotation angle $\omega$:
\begin{equation}
    \sigma^2 = \int \omega^2 \rho(g) d\Omega 
\end{equation}
 

\section{Acknowledgements}
 We would like to acknowledge Moritz Stammer for helpful discussions and the ESRF ID13 staff for supplying beamtime and support during the experiment. This work is funded by the European Union, European Research Council H2020, TexTOM (grant no. 101041871). Views and opinions expressed are those of the author(s) only and do not necessarily reflect those of the European Union or the European Research Council. Neither the European Union nor the granting authority can be held responsible for them.




\bibliographystyle{ieeetr}

\bibliography{iucr}





\end{document}